\documentclass[a4paper, amsfonts, amssymb, amsmath, reprint, showkeys, nofootinbib, twoside]{revtex4-2}
\usepackage[english]{babel}
\usepackage[utf8]{inputenc}
\usepackage[colorinlistoftodos, color=green!40, prependcaption]{todonotes}

\usepackage[pdftex, pdftitle={Article}, pdfauthor={Author}]{hyperref} 
\usepackage{amsthm}
\usepackage{mathtools}
\usepackage{physics}
\usepackage{xcolor}
\usepackage{graphicx}
\usepackage[left=23mm,right=13mm,top=35mm,columnsep=15pt]{geometry} 
\usepackage{adjustbox}
\usepackage{placeins}
\usepackage[T1]{fontenc}
\usepackage{lipsum}
\usepackage{ulem}
\usepackage{csquotes}
\usepackage{float}
\usepackage[section]{}

\begin{document}%
    \title{Symbiotic Brain-Machine Drawing via Visual Brain-Computer Interfaces}
    \author{Gao Wang$^{1}$, Yingying Huang$^{1,2}$, Lars Muckli$^2$, Daniele Faccio$^{1}$}
    \email[Correspondence email address: ]{daniele.faccio@glasgow.ac.uk}
    \affiliation{$^{1}$School of Physics \& Astronomy, University of Glasgow, Glasgow, UK, G12 8QQ.\\
    $^{2}$School of Psychology and Neuroscience, University of Glasgow, Glasgow G12 8QB, UK}

    \date{\today}

    \begin{abstract}
    Brain-computer interfaces (BCIs) are evolving from research prototypes into clinical, assistive, and performance enhancement technologies. Despite the rapid rise and promise of implantable technologies, there is a need for better and more capable wearable and non-invasive approaches whilst also minimising hardware requirements.  We present a non-invasive BCI for mind-drawing that iteratively infers a subject's internal visual intent by adaptively presenting visual stimuli (probes) on a screen encoded at different flicker-frequencies and analyses the steady-state visual evoked potentials (SSVEPs). A Gabor-inspired or machine-learned policies dynamically update the spatial placement of the visual probes on the screen to explore the image space and reconstruct simple imagined shapes within approximately two minutes or less using just single-channel EEG data. Additionally, by leveraging stable diffusion models, reconstructed mental images can be transformed into realistic and detailed visual representations. Whilst we expect that similar results might be achievable with e.g. eye-tracking techniques, our work shows that symbiotic human-AI interaction can significantly increase BCI bit-rates by more than a factor 5x, providing a platform for future development of AI-augmented BCI.
    \end{abstract}

    \keywords{SSVEP, Human-machine collaboration, Brain-reading, Gabor filter, Brain-computer interface}

    \maketitle

     \section{Introduction}
     
     Brain-computer interface (BCI) technologies enable direct communication between the brain and a computer, with applications ranging from controlling the computer as a means to regain, for example, the ability to move or navigate the world, to the actual decoding of human thought \cite{abdulkaderBrainComputerInterfacing2015, biEEGBasedBrainControlledMobile2013, chenHighspeedSpellingNoninvasive2015, fouadBrainComputerInterface2015,lelievreSingleTrialBCI2013}. BCIs can be broadly divided into two paradigms. Passive BCIs infer cognitive states such as workload or fatigue, from spontaneous brain activity  \cite{zanderPassiveBrainComputer2011, alimardaniPassiveBrainComputerInterfaces2020}, while active BCIs rely on subjects intentionally modulating their neural signals to issue commands  \cite{mladenovicActiveInferenceUnifying2020,norciaSteadystateVisualEvoked2015}. \\
     The ability to directly observe and decode subjective visual imagery is of fundamental importance, holding transformative potential for assistive communication, creative co-design and the study of mental health. The central limitation of existing methods is their reliance on pre-trained, data-hungry models, which act as a "dictionary" of known images {\cite{benchetritBrainDecodingRealtime2024a, duFMRIBrainDecoding2022a, liuBrainCLIPBridgingBrain2023, wangMindBridgeCrossSubjectBrain2024}}. This fundamentally restricts the creative potential of mind drawing.

    Following the development of a “P300 speller”-based painting system \cite{zicklerBrainPaintingUsability2013}, a BCI painting system employing a hybrid SSVEP/P300 control approach was later proposed \cite{tangBCIPaintingSystem2022}. However, both methods rely heavily on using the SSVEP to "click" the static graphical subject interfaces (GUIs), and they lack the intelligence to predict the subject's intention {\cite{mullerEffectsSpatialSelective1998}}.
    
     Noninvasive neural decoding methods have been developed using the main brain sensing technologies, i.e. electroencephalography (EEG), functional magnetic resonance imaging (fMRI), functional near-infrared spectroscopy (fNIRS), and magnetoencephalography (MEG). 
     In recent years, researchers have implemented neural decoding methods to monitor the brain state or to communicate with others \cite{nicolas-alonsoBrainComputerInterfaces2012, norciaSteadystateVisualEvoked2015, wangComputationalGhostImaging2023, wangHumancentredPhysicalNeuromorphics2024} but also to reconstruct mental or visual images from brain signals. 
     For example fMRI, with its high spatial resolution, has provided foundational insights into how visual information is encoded in the brain, has been used to reconstruct images from blood-oxygen-level-dependent (BOLD) signals \cite{miyawakiVisualImageReconstruction2008,duFMRIBrainDecoding2022, horikawaAttentionallyModulatedSubjective2020, leBrain2PixFullyConvolutional2021, mengDualGuidedBrainDiffusion2023, seeligerGenerativeAdversarialNetworks2018, shenDeepImageReconstruction2019}. \\
     Recent work has also highlighted the possibiliy to decode visual imagery via fNIRS \cite{adamicProgressDecodingVisual2024}.
     MEG  was proposed for real-time reconstruction of visual perception \cite{benchetritBrainDecodingRealtime2024} and classification \cite{vandenieuwenhuijzenMEGbasedDecodingSpatiotemporal2013}.
     EEG is more portable than MEG, (the fewer EEG channels, the more portable) and also has high temporal resolution and has been implemented to achieve reconstruction, from visual texture to natural images, with the help of Neural Networks \cite{khaleghiVisualSaliencyImage2022, liVisualDecodingReconstruction2024, panReconstructingVisualStimulus2024, singhEEG2IMAGEImageReconstruction2023, sokacBridgingArtificialIntelligence2024a, wakitaPhotorealisticReconstructionVisual2021, benchetritBrainDecodingRealtime2024, chenSeeingBrainConditional2022, guentherImageClassificationReconstruction2024, leeExploringAbilityClassify2022, shimizuImprovingClassificationReconstruction2022}. More recently, the `BrainVis' approach \cite{fuBrainVisExploringBridge2025} demonstrated state-of-the-art semantic fidelity reconstructions and generation quality. These mind-drawing or image-reconstruction approaches rely on high-dimensional subject data and deep learning algorithms, which are complex to implement and typically do not generalise to unseen imagery.
     
     Despite significant progress in the field, achieving true mind-drawing in real-time with model-free and high-fidelity reconstruction of arbitrary images from imagination alone remains a challenge. A further opne challenge for the field is relatively low information transfer bit-rates of BCIs, with hints that these might be fundamentally limited to $\sim10$ bits/second \cite{zhengUnbearableSlownessBeing2025}.

     Here, we propose a non-invasive neural decoding framework that overcomes key limitations of existing mind-drawing approaches. We develop reconstruction approaches based on an iterative collaborative search in image space, guided by dynamically updated sampling weights for successive visual stimuli (probes) that are presented to the subject. Crucially, our system does not rely on any pre-trained generative models. Instead, it generates the layout of subsequent visual probes in real time through a Gabor analysis of accumulated neural evidence from steady-state visual evoked potentials (SSVEPs). This allows the system to adaptively adjust its detection strategy based on the structure of mental imagery as this iteratively emerges during the drawing process, efficiently focusing resources on the most informative regions in the visual field.

     
     Preliminary human experiments demonstrate that this neuro-adaptive computational imaging framework can reconstruct simple, imagined visual shapes in approximately two minutes or less using only single-channel SSVEP data. The bit-rate of the mind-drawing process is improved by more than up to a factor 5x when compared to a simple readout of the SSVEP signal from the same device.
    
    \section{Results}
\begin{figure}[!t]
    \centering
    \includegraphics[width=\linewidth]{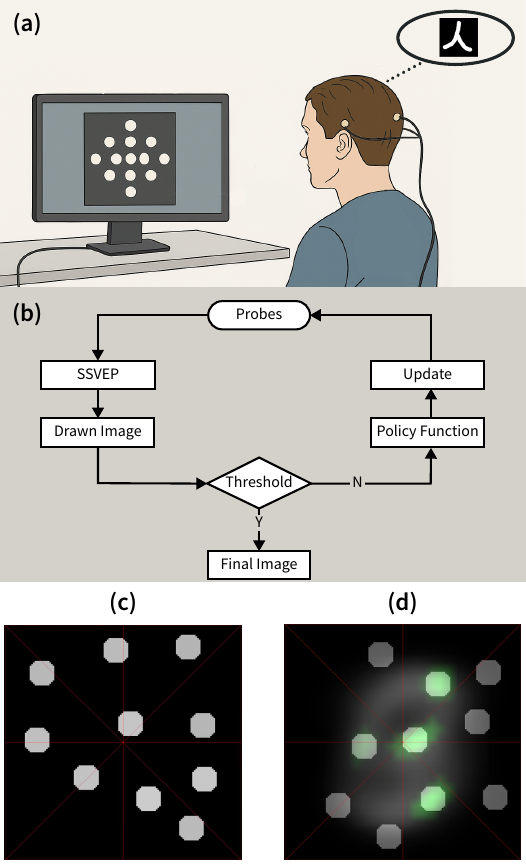}\\
    \caption{Experimental setup and workflow. 
    (a) Setup of the EEG-based mind-drawing system. The EEG device is custom-built and consists of a headband housing three wet electrodes (using saline solution to improve contact): two placed on the temples (ground and reference) and one at the occipital Oz position. The subject selects the disc that overlaps most with their imagined image. 
    (b) Workflow of the mind-drawing process (Y: Yes; N: No). 
    (c) Example screen-shot stimulus under the Gabor policy function, where discs are randomly arranged and each flickers at a unique frequency. 
    (d) Example screen-shot of the stimulus under the data-driven policy function, where discs are randomly arranged, with some representing green features and others appearing as standalone elements.}
    \label{fig:setup}
    \end{figure}

    The experimental setup is illustrated in Fig.~\ref{fig:setup}(a). The subject was seated in front of a computer monitor while wearing a single-channel EEG device and was instructed to select the flickering disc that overlapped most with their imagined image. Figure~\ref{fig:setup}(b) illustrates the workflow of mind-drawing where the human is in the loop throughout the process. The disc positions are updated by the policy function based on the accumulated historical data.
    At the beginning of each iteration, 10 white visual stimulus probes (flickering discs) were displayed on a {black-background} screen (as shown in Fig.~\ref{fig:setup}(c)). The subject was then instructed to focus on the region where the overlap between one presented probe and the mentally imagined object appeared to be strongest within each iteration. After 4 seconds of flashing stimulation, the SSVEP response was recorded and the Canonical Correlation Analysis (CCA) value (a correlation coefficient ranging from 0 to 1 \cite{suhendraCanonicalCorrelationAnalysis2024}) was used to weight the selected probe/pattern that was then added to the drawn image as real-time EEG feedback to the subject displayed on the screen background in red. This process was repeated iteratively, with updated probe positions presented to the subject each time. The updated positions were determined by a policy function that determined the placement of the next set of probes (see Methods for a detailed description of the policies and how these were applied). We have tested two types of policy functions. Figure~\ref{fig:setup}(c) shows an example stimulus under the Gabor policy function, in which discs are randomly arranged and each disc flickers at a unique frequency. Figure~\ref{fig:setup}(d) illustrates an example stimulus under the data-driven policy function, where disks that have been selected by the subject at the previous iterations actually now represent machine-learned 'basis functions', shown in green.  The final image therefore appears from the superposition of multiple basis functions as opposed to the simple discs used in the Gabor analysis approach (see details below and in Methods). 
    
    The selected probes were iteratively updated and added to the drawn-image. In our tests, we found that the optimal choice was a total of 25 iteration runs for the Gabor policy and seven iterations for the more efficient choice (but more constrained, see below) data-driven policy case.  A 25-session measurement lasts 2.5 minutes, also including 2-second resting periods. The final drawn image $I_n(x,y)$ was then reconstructed based on the accumulated weighted pattern, 
    \begin{equation}
        \label{eq:I_cumsum}
        \begin{aligned}
            I_n(x,y) & = \sum_{i=1}^{n} B_{j} P_j(x,y).
        \end{aligned}
    \end{equation}
    where {$B_{j}$ is the weight, and }the pattern $P_j(x,y)$ is a Gaussian disc in the Gabor policy or is a 'basis pattern' in the data-driven policy (see Methods for a detailed description). Through iterative refinement guided by real-time EEG feedback and the selected policy function, the reconstructed image gradually converged toward the imagined object by the participant.

    We conducted three single-channel EEG-based mind-drawing experiments:\\
    \textbf{Experiment 1}: Gabor-policy inspired mind drawing, including eight subjects with each subject drawing 3 different images, so to achieve multi-subject validation and estimate similarity and information rate analysis; \\
    \textbf{Experiment 2}: Data-driven mind drawing, including one subject for different handwritten digit images, demonstrates the effect of data-driven policy as characterized by mutual information;\\
    \textbf{Experiment 3}: Gabor-policy mind drawing enhanced with stable diffusion, including different drawing results under the same prompt from two different subjects, leading to different detailed images from SD's output. 
    
    \subsection{Gabor-inspired Mind Drawing}
\begin{figure}[!t]
        \centering
        \includegraphics[width=\linewidth]{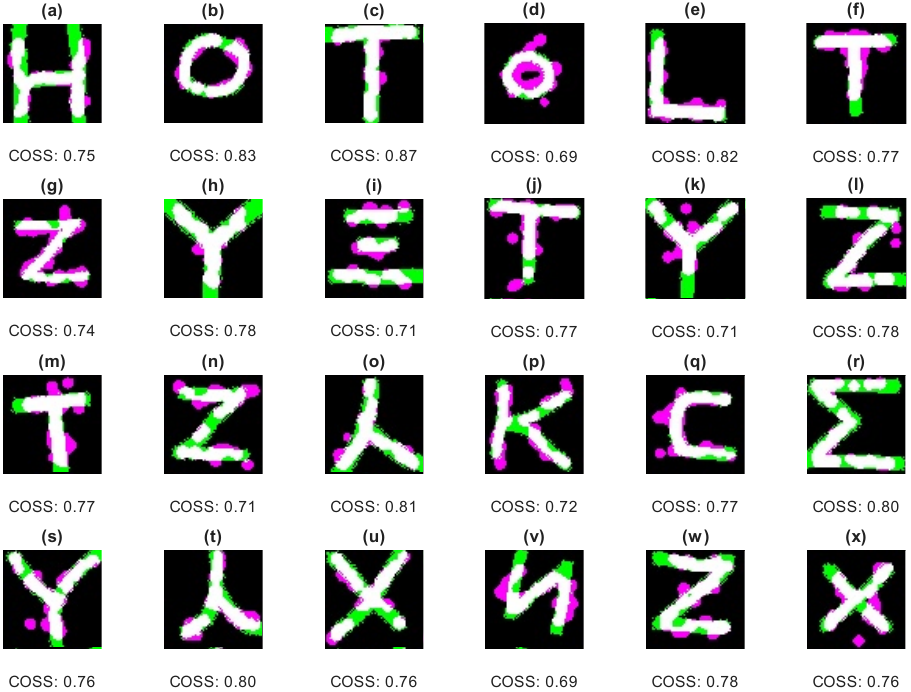}
        \caption{Imaging results from eight subjects, each targeting three imagined shapes. Panels are grouped three at a time, one group for each subject, i.e. panels (a–c) correspond to the first subject, (d–f) to the second subject, and so forth. 
        Green-colored images represent the handwritten target after image resizing, while pink-colored images indicate the reconstructed images generated by our BCI system. Overlapped regions highlight the common areas between the two images, indicating reconstruction accuracy. 
        We calculated the cosine similarity (COSS) between each ground-truth and reconstructed image pair to quantify accuracy. The overall average COSS across all subjects was 0.76 (± 0.04), indicating a good level of reconstruction fidelity.}
        \label{fig:8_subjects}
    \end{figure}

    To establish a ground truth, subjects were first required to hand-draw the image they intend to imagine in the experiment and then perform the mind-drawing task using the proposed system. We evaluated the performance across multiple subjects to assess the generalizability and robustness of the approach. Eight healthy human subjects participated and each completed mind-drawings of three different images with 25 iterations. The results are illustrated in Fig.~\ref{fig:8_subjects}, showcasing the system's ability to reconstruct simple geometric shapes with varying degrees of accuracy across subjects. Green-colored images represent the handwritten target after a shape-preserving transformation (i.e. rescaling, rotation, translation) so that it has same size as the mind-drawn image, shown as a pink-colored image. The overlap regions between the ground-truth and the reconstructed images appear in white. We calculated the cosine similarity (COSS) between each ground-truth and reconstructed image pair to quantify accuracy. The overall average COSS across all subjects was 0.76 (± 0.04), indicating a good level of reconstruction fidelity.

\begin{figure}[!t]
        \centering
        \includegraphics[width=\linewidth]{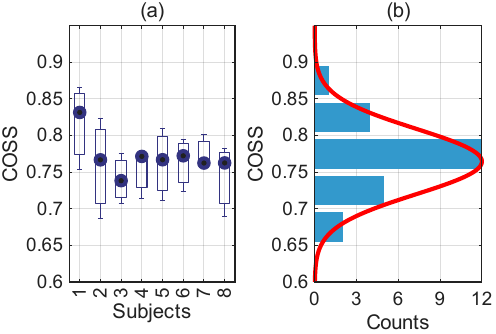}
        \caption{Distribution of image cosine similarity (COSS) values between target and reconstructed image pairs. (a) Box plot of COSS values for each subject.  
                (b) Histogram of COSS values across all subjects.}
        \label{fig:boxplot}
    \end{figure}

   Figure~\ref{fig:boxplot} shows a COSS box plot and histogram distributions that illustrates how the COSS values for each subject are relatively consistent, with most values ranging between 0.7 and 0.8. This indicates that the system is robust and generalises well across individuals.  
    The box plot indicates that there is some variability in reconstruction accuracy between subjects, with some achieving higher COSS values than others. This variability may be attributed to individual differences in mental imagery, attention and focus during the task.  
    %

\begin{figure}[!t]
        \centering
        \includegraphics[width=.8\linewidth]{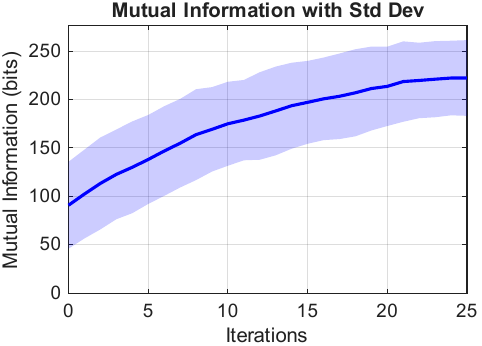}
        \caption{Mutual information (with standard deviation across iterations and subjects) increases with the number of iterations, although the rate of increase slows down as more iterations are added. The final MI reaches 222~bits starting from 91~bits (MI from an image with all black pixels), corresponding to an average rate of 1.31~bit/s.
        }
        \label{fig:MI_Plot}    
    \end{figure}

    We then estimated the mutual information (MI) between the target and the reconstructed images, as MI captures the amount of shared information and provides an information-theoretic measure of reconstruction fidelity. The MI (see Methods, Eq.~\eqref{eq:MI}) results for our measurements are shown in Fig.~\ref{fig:MI_Plot}: the plot starts at 91 bits (corresponding to a full black image), and the MI of all subjects increases with the number of iterations and after 25 iterations reaches 222 bits, corresponding to a bit rate of 1.31 bits/second. 
    %
    We estimated the more widely used Information Transfer Rate (ITR see Methods, Eq.~\eqref{eq:ITR_BCI}), which provides the maximum theoretical information transfer rate (see Methods, Eq.~\eqref{eq:ITR_max}), $\mathrm{ITR}_{\max} \approx 0.83~\text{bit/s}$.
    
    Remarkably, we find that the measured bit rate based on the mutual information between the target and reconstructed images (1.31 bits/second) is $\sim1.6$x larger than the maximum ITR predicted for a standard BCI (0.83 bits/second) {\cite{wolpawEEGbasedCommunicationImproved1998a,sadeghiAccurateEstimationInformation2019}}. We attribute this to the effect of the iterative feedback and optimisation policy that is used to choose the optimal probe placement at each iteration, which in turn is based on information from the previous iteration. This indicates that our real-time brain-computer cooperation (implemented through our iterative optimisation policy) leads to a much higher bit rate and higher performance compared to what one might expect from a simple, one-way communication BCI. 

    \subsection{Data-driven Mind Drawing}

    To further accelerate and enhance reconstruction, we leveraged prior datasets to inform probe selection, specifically by using machine-learning to create 'basis functions' that are used in place of the Gaussian discs (used in the Gabor policy case) to iteratively create the image. We therefore compromise image generalisation in favour of reconstruction speed as the use of prior datasets (that determine the specific shapes of the 'basis functions') unavoidably implies that the images that are drawn will need to resemble images in the dataset. Specifically, we conducted the second experiment with the same procedures as in Experiment 1 but using the MNIST dataset of handwritten digits with a data-driven policy function (see Methods).
    The system successfully reconstructed digits based on SSVEP signals, demonstrating its ability to handle more complex visual stimuli beyond simple geometric shapes although, of course, these stimuli do need to bear resemblance to the images in the data training set. Reconstruction accuracy varied among digits, with structurally simpler digits generally yielding higher fidelity.
    
\begin{figure}[!t]
        \centering
        \includegraphics[width=\linewidth]{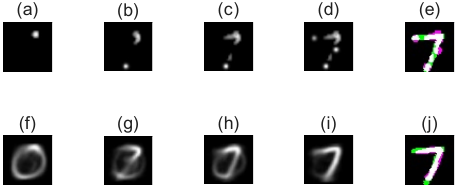}
        \caption{Reconstruction results for the MNIST digit “7.” 
        The first row (a–d) shows the step-by-step reconstruction of the digit  from SSVEP signals by placing a Gaussian disc at the location of each disc selected by the SSVEP signal. 
        The second row (f–i) presents the corresponding images generated by the system at each step. 
        For simplicity, we only show the images every 2 iteration steps.
        Panels (e) and (j) overlay the binarised reconstructed and predicted images with the aligned handwritten target in different colour channels: green indicates the aligned handwritten target, pink represents the reconstructed mental image, and overlapped regions highlight areas of agreement, demonstrating reconstruction accuracy. 
        Grey regions indicate agreement, while colored regions highlight intensity differences.
        The COSS values for the binarised reconstructed and predicted images are 0.75 and 0.78, respectively
        , indicating a high degree of similarity to the target image.}
        \label{fig:MNIST_reconstruction}    
    \end{figure}

    As shown in Fig.~\ref{fig:setup}(d), selected discs are shown with the corresponding 'basis function' in green in the background.
    Figure~\ref{fig:MNIST_reconstruction} presents representative results for the digit “7” with increasing iterations from left to right. The top row shows, for comparison, the `raw' image that is obtained based on Eq.\eqref{eq:I_cumsum} and showing a Gaussian disc (similarly to what was done for the Gabor policy approach) at the position of the disc from the SSVPE signal. The bottom row presents the data-driven policy image results, derived from the same process but now using the weighted `basis functions' and near-neighbour analysis. As can be seen from the last panels in each row (h) and (p), respectively, the full data-driven pipeline provides images that are better representations of the intended image (i.e. the panels show larger `white shaded overlap regions' in (p) compared to (h) ) As more 'basis functions' are accumulated, the reconstructed mental image becomes progressively more detailed and structured. 

   The detail with which the image generation occurs is also interesting - at the early stages, the reconstruction is ambiguous, sometimes resembling a “0”. With additional iterations, the image gradually clarifies, oscillating between “7” and “3,” and ultimately converging to a clear “7” with high fidelity. Notably, in this experiment, the system requires only five iterations to produce a rough outline of the target digit, and six iterations to achieve a recognisable form. Each subsequent iteration further refines the reconstruction, illustrating the system’s capacity for adaptive learning and progressive improvement based on accumulated SSVEP evidence. Additional representative results are provided in the SM.

\begin{figure}[t]
        \centering
        \includegraphics[width=.8\linewidth]{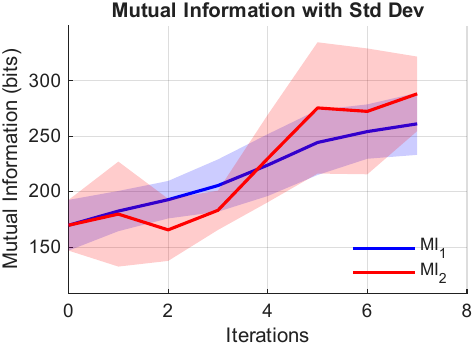}
        \caption{Mutual information during the reconstruction of MNIST digits, averaged over four different digit reconstructions (``7'', ``2'', ``4'', and ``8''). $\text{MI}_1$ and $\text{MI}_2$ denote the mutual information of the SSVEP reconstruction alone (i.e. same approach as top row in Fig.~\ref{fig:MNIST_reconstruction})  and the data drvien policy images  (i.e. same approach as bottom row in Fig.~\ref{fig:MNIST_reconstruction}), increasing from 170 bits to 261 bits and 288 bits, respectively. 
        The corresponding bit rates are 3.25 bits/s and 4.21 bits/s. }
        \label{fig:MI_MNIST_Plot}    
    \end{figure}

    We estimated the MI between the target and reconstructed images, as done in  Experiment~1. As shown in Fig.~\ref{fig:MI_MNIST_Plot}, the MI values demonstrate the progressive improvement of reconstruction quality across iterations, reaching rates as high as 4.21 bits/s. This increase in information rate is driven by the machine-learned data-driven policy which is introducing additional information at each iteration. We need however, to recall that this increase of bit rate from 0.83 to 4.21 bit/s (an increase ofr more than a factor 5x) comes at the expense of generalisability. 
    \subsection{Stable Diffusion enhanced mind drawing}
    
\begin{figure*}[t]
        \centering
        \includegraphics[width=\linewidth]{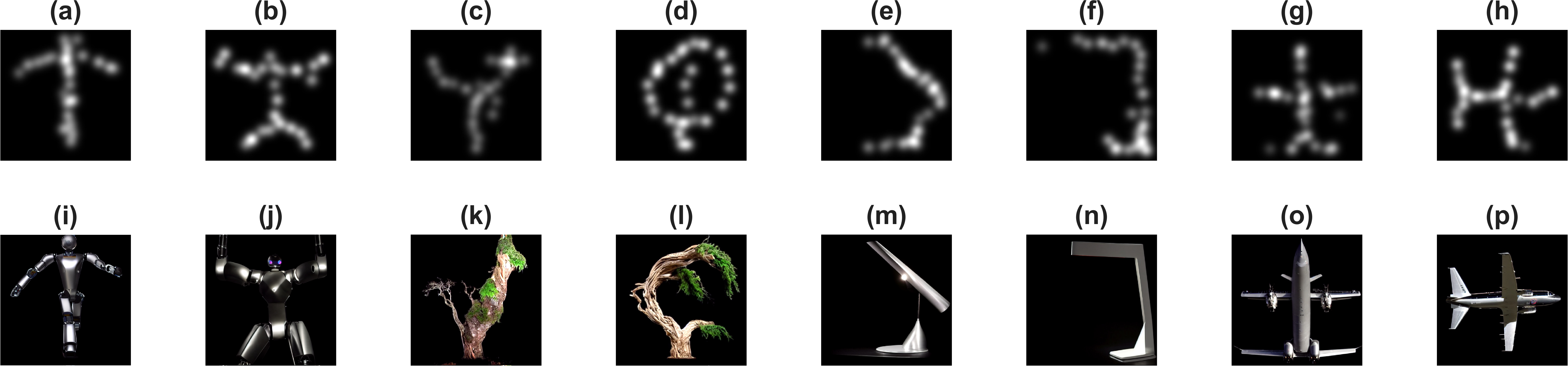 }
        \caption{Examples of mind drawing with the gabor policy approach that is enhanced with a final  Stable Diffusion step. Panels~(a–h) show examples of  Gabor policy drawings. Panels~(i–p) show the corresponding final images  generated under the same SD text prompt (these are grouped into 4 pairs, i.e. (i)-(j), (k)-(l), (m)-(n), (o)-(p) have the same SD text prompts). Prompt and model details are provided in the SM.}
        \label{fig:SD_example2}
    \end{figure*}

    To evaluate whether the reconstructed images could be further refined into realistic visual representations, we use the same approach as in Experiment 1, i.e. we adopt the Gabor-policy approach but then augment this with a final step that  applies Stable Diffusion as a generative enhancement model. Stable Diffusion is a generative model capable of producing high-quality images from text prompts and/or image inputs. The rationale for this was to test whether conditioning the evolving mental image would allow Stable Diffusion to complete and sharpen the reconstructions, driven by neural signals.
    {As shown in Fig.~\ref{fig:SD_example2}, 4 different classes of images are illustrated, including a robot, a tree, a desk lamp, and an aircraft. For each class we show two mind-drawing results under the same prompt (but different mind-drawing sessions) to recreate a new detailed image from a mind-drawn image.} 
    This approach bridges the gap between abstract neural reconstructions and photorealistic imagery, enabling applications such as creative co-design, assistive communication, and personalised content generation.

    \section{Discussion and conclusions}
  We propose a non-invasive neural decoding approach that enables mind-drawing through an adaptive, iterative process that is based on a straightforward, single-channel EEG device.

    The system's performance across multiple subjects demonstrates its robustness and generalizability, with promising results in reconstructing simple visual forms. 
    Interestingly, we found that the mutual information of the mind-drawing system is significanlty larger than the maximum expected ITR. This is a result of the iterative and policy-driven optimisation process of the computer prompts, supporting the concept of true brain-machine cooperation. 

 { We note that a similar approach could in principle also be implemented with other technologies, specifically for example with an eye tracker. However, the key point we wanted to make here is that it is possible to achieve this with EEG or neural signals alone as this might be relevant for future development towards more complex mind-driven systems. Our approach also has several advantages. First, it is a very  affordable system that can be home-built for less than \$100. Second, it can use SSVEP-based attention as a graded output to weight patterns (the weighting is derived directly form the intensity of the SSVEP signal), whereas an eye tracker typically provides only a binary focus position. Finally, SSVEPs can be modulated both by attention and by the focus point, while eye trackers rely solely on eye movements.}

    The integration of Stable Diffusion further enhances the system's capabilities, transforming abstract neural reconstructions into realistic and detailed visual representations. This bridges the gap between mental imagery and photorealistic imagery, enabling applications such as creative co-design and personalised content generation.

    The ability to infer non-verbal, open-ended spatial geometric content in real-time could provide an opportunity for new applications in neuroscience, cognitive science, and assistive communication. By prioritising subject experience, the system allows for intuitive interaction and adaptation to individual mental imagery patterns. We expect that with further developments, by focusing on overlapping regions between visual probes and imagined objects such that the system effectively captures the subject's intent, will also enable a more natural and engaging BCI experience.


    \section{Methods}\label{Section: Method}

    \subsection{EEG System and Pattern Stimulus}

    We used a three-electrode EEG device, with one active electrode placed at the Oz position to capture the SSVEP signal from the primary visual cortex, a reference electrode above the left ear (M1 position), and a ground electrode above the right ear (M2 position). Participants were seated approximately 50–70 cm from the computer screen. The system used MATLAB for real-time data acquisition and analysis, with visual stimuli presented using Psychtoolbox (PTB). EEG signals were sampled at 1 kHz via a microphone-based ADC sound card (Startech C-Media, 361 ICUSBAUDIO). The setup was implemented on a Windows 10 desktop equipped with a 480Hz 26.5-inch QHD (2560 x 1440) monitor (ROG Swift OLED PG27AQDP), working at 120 Hz or 240 Hz to ensure stable visual presentation. A square stimulus area of $1440 \times 1440$ pixels was centred on the screen.

    At the beginning of the experiment, each subject sketched a simple imagined shape, such as a letter Y, on the canvas (details in SM) using a mouse. This image served as the target for the subsequent mental imagery task. After completion of the drawing and before the periodic flashing disc stimulus, a static image containing 10 discs was displayed on the screen shortly, and then start flashing.  {The number of discs on the screen is determined by two factors: (1) the size of each disc (a 3 degree visual angle provides a robust SSVPE response) and their spacing (the centre-to-centre spacing is kept 2x the stimulus size i.e. the centre-to-centre spacing is 300 pixels.)(2) the flicker frequency of each diske is chosen within the 10–19 Hz range and with a separaton of 1 Hz.}

    After sketching the image using a drawing board with a mouse,  subjects were informed to use mind drawing to draw the outline first and then go into detail, following the following strategy to enhance reconstruction accuracy and speed:
    \begin{enumerate}
        \item \textbf{Focus on Overlapping Areas:} Select the disc with the largest intersection or overlap with the imagined object. Concentrate on the overlapping region of the disc rather than its centre.
        \item \textbf{Prioritise Outline Information:} Begin by marking points with larger spacing i.e. attempt to first reconstruct the large-scale features and basic structure of the object. For instance, place points on the key strokes of the object (e.g., for the letter "Z," mark points on the horizontal, diagonal, and vertical strokes).
        \item \textbf{Refine Details:} Gradually select points with smaller spacing to add finer details and progressively refine the shape of the object.
    \end{enumerate}
    %

    In the Gabor policy function mode, as shown in shown Fig~\ref{fig:setup}(c), the stimulus consists of 10 discs, each flickering at a distinct frequency. These discs are placed at random locations on the screen, with each disc serving as a potential probe location for the subject’s imagined image. 


    {In the Data-driven policy function mode, as shown in Fig.~\ref{fig:setup}(d), the stimulus also comprises 10 discs, with some representing green features (`basis function' patterns from the MNIST dataset; details in SM) and others serving as standalone elements. The subject is instructed to select the disc whose presented pattern overlaps most closely with their imagined image. 
Additionally, a reconstructed image is displayed in the background throughout the experiment. Initially, this background corresponds to the dataset’s average image, but it is iteratively updated based on the subject’s selections and the system’s ongoing reconstruction. This background serves as visual feedback and context, allowing the subject to monitor the progression of mental image reconstruction.}
    

    
    The CCA algorithm  was applied to determine which probe the participant was focusing on. The participants focused on the disc that overlapped the most with their imagined image, eliciting an SSVEP response in the EEG signal. In the next iteration, the selected disc was weighted with its corresponding CCA value and displayed as part of the red background in the static image at the beginning of the subsequent iteration.

    \subsection{Gabor-inspired Policy Function}
    Gabor filters are specialised bandpass filters sensitive to textures and edges in specific directions and frequencies. A set of Gabor filters with varying orientations and scales acts as a detector array for "texture and structure." This is particularly relevant because studies have shown that the receptive fields  in the primary visual cortex (V1) closely resemble Gabor functions. Thus, employing Gabor analysis to process subject-drawn images is chosen so as to emulate the early stages of human visual processing \cite{huangGaborNonGaborNeural2023}.
    
    We apply a 2-step Gabor analysis (including feature kernel and probability kernel, detailed in SM) to the current estimated image, $I_{\text{pt}}$, which is composed of a sparse set of discrete CCA-weighted pixel points (centre of the disc), as shown in Eq.~\eqref{Gabor_update}.
    
    \textbf{1. Extraction of Gabor features:} 
    Gaussian smoothing, $\sigma = 50$ pixels, is applied to $I_{\text{pt}}$, producing a blurred image $I_{\text{gs}}$ highlighting low-frequency structure. {Here, $I_{\text{gs}}$ is equal to $I_n$ in Eq.~\eqref{eq:I_cumsum}}. The Gabor feature kernel, $G_{\text{feat}}$, is convolved with $I_{\text{gs}}$ to extract initial structural features, $I_{gg}$, such as edges and stripes.  We then subtract the mean of each filter channel, $\bar{I}_{gg}$, and the result is passed through a ReLU function (to retain only the positive components) followed by weighting of the original point-based estimate $I_{\text{pt}}$, yielding a weighted feature map $I_{\text{feat}}$.
    
    \textbf{2. Sampling weights:}
    This feature map, $I_{\text{feat}}$, is subsequently convolved with the Gabor probability kernel,  $G_{\text{prob}}$, producing the final probability map $I_{\text{prob}}$ (also referred to as a sampling weight), which guides the spatial placement of probe disks in the next iteration. $G_{\text{feat}}$, $G_{\text{prob}}$, and one example of this 2-step flow is detailed in SM. Summarising the steps above in formulas, we have:
    \begin{equation}
        \label{Gabor_update}
        \begin{aligned}
            I_{\text{gs}}   & = I_{\text{pt}}\ast \text{Gaussian},                                              \\
            I_{\text{gg}}   & =I_{\text{gs}}\ast G_{\text{feat}},\\
            I_{\text{feat}} & = I_{\text{pt}}\cdot \text{ReLU}\left( I_{\text{gg}}- \bar{I}_{\text{gg}}\right), \\
            I_{\text{prob}} & = I_{\text{feat}}\ast G_{\text{prob}}.
        \end{aligned}
    \end{equation}
    If the class of images is limited to a specific set, then as shown in the next section, a Data-driven Policy Function can be used to enhance the reconstruction process further. In this case, the Gabor feature analysis is replaced by a data-driven approach that leverages prior knowledge from a dataset, such as MNIST for handwritten digits. This allows the system to adaptively select probes based on the current estimate of the mental image.

    \subsection{Data-driven Policy Function}\label{section:Data-driven Policy Function}
    The data-driven policy function is designed to accelerate the reconstruction process by leveraging prior knowledge from a dataset, such as MNIST for handwritten digits. 

    In our experiments, we use the MNIST dataset, which contains 60,000 training and 10,000 test images of handwritten digits (each $28 \times 28$ pixels). Non-negative Matrix Factorisation (NNMF) decomposes these images into a set of basis patterns (atoms) and their corresponding weights. This allows images to be represented in a lower-dimensional, more efficient feature space for reconstruction. Here,  we select 25 atoms from the NNMF decomposition to serve as the feature space for image representation and reconstruction, as detailed in the SM.

    In our framework, atom/basis patterns are presented as probe options, similar to the previous approach of displaying 10 disc probes. As shown in Fig.~\ref{fig:setup}(d), some probes correspond to basis patterns (highlighted with a green feature background), while others correspond to discs themselves.  The participatn is asked to prioritize selection of an atom pattern. If no atom pattern overlaps with the subject’s imagined image, the subject is instructed to select a normal probe (without green feature background). If there is still no overlap for any of the discs, the subject focuses on a neutral area.
    
    The `seed image' at the beginning of the experiment is taken simply as the average image of the whole dataset and is projected into the atom feature latent space as a vector, guiding the selection of the feature patterns. 


    To guide the selection of the next atom pattern, we first identify the 100 nearest MNIST images to current reconstucted image $I_{n}$ in the latent space. The average of their latent vectors yields a $(1 \times 25)$ vector $W_{n}$, which is then used as the sampling weight for the next atom probe selection. Moreover, the prediction of the subject’s mental image, obtained  from $W_{n}$ is displayed as a red background on the screen, visible to the subject. 

    
    \subsection{Information Rate Calculation}
    For a single pixel, the mutual information can be expressed as
    \begin{equation}\label{eq:MI}
    I(X;Y) = p \log_2 (2p) + (1-p) \log_2 \bigl[ 2(1-p) \bigr],
    \end{equation}
    where $p$ denotes the probability of a correct match rate between two images.
    If $n$ is the number of image pixels, then the mutual information for the entire image can be estimated as
    $\mathrm{MI} = n \, I(X;Y)$. In our experiment, $p$ is estimated by comparing the whole image with the ground truth and calculating the correct rate for pixels.
    
     We can also estimate the maximum theoretical information transfer rate for a BCI using the standard equation (the Wolpaw formula) {\cite{wolpawEEGbasedCommunicationImproved1998a}},
    \begin{equation}\label{eq:ITR_BCI}
        \mathrm{ITR} = \frac{1}{T} \left( \log_{2} N + P \log_{2} P + (1-P) \log_{2} \frac{1-P}{N-1} \right)
    \end{equation}
    where
    $N$ is the number of targets, $P$ is the probability of correct target identification (i.e., accuracy), and $T$ is the average time required for a single selection. 
    In our experiments, $N=10$, $T=4$ seconds (the time for a single iteration during which a single target is identified with probability $P=1$), and the maximum theoretical ITR is:
    \begin{equation}\label{eq:ITR_max}
        \mathrm{ITR}_{\max} = \frac{1}{T} \log_{2} N \approx 0.83 ~ bit/s
    \end{equation} 
    %
    
    \section*{Acknowledgments}
    G.W. and Y.H. acknowledge the support of the China Scholarship Council (CSC). D.F. acknowledges support from the Royal Academy of Engineering Chair in Emerging Technologies programme and the UK Engineering and Physical Sciences Research Council (grant EP/T00097X/).

    \section*{Conflicts of Interest}
    All authors declare that there is no conflict of interest.
    \vspace{1em}
    
    \bibliographystyle{naturemag}
    \bibliography{References_.bib}
\end{document}


\title{ Symbiotic Brain-Machine Drawing via Visual Brain-Computer Interfaces:
Supplementary Material}

\author{Gao Wang$^{1}$, Yingying Huang$^{1,2}$, Lars Muckli$^2$, Daniele Faccio$^{1}$}
    \email[Correspondence email address: ]{daniele.faccio@glasgow.ac.uk}
    \affiliation{$^{1}$School of Physics \& Astronomy, University of Glasgow, Glasgow, UK, G12 8QQ.\\
    $^{2}$School of Psychology and Neuroscience, University of Glasgow, Glasgow G12 8QB, UK}

\date{\today}

\begin{abstract}
Supplementary Material with additional theoretical descriptions and experimental results.
\end{abstract}


\maketitle



\subsection{Gabor-inspired Policy} 

We apply a set of multi-directional Gabor filters to the user's drawing trajectory images. Given the strong resemblance between Gabor functions and the receptive fields of neurons in the primary visual cortex (V1), this analysis quantitatively captures the visual structure of the user's creation. By identifying the Gabor filter orientation with the strongest response, the engine can dynamically infer the dominant geometric trend of the current drawing. 
For example, a pronounced response in the vertical direct filter indicates a strong vertical trend in the user's creation. The inferred probability distribution is then used to adaptively position candidate targets (flashing disks) along these high-probability paths, providing real-time assistance aligned with the user's spatial intention. Specifically, many objects, such as letters (“F”) or symbols (a cross), are composed of lines or curves. 

\begin{figure}[htbp]
    \centering
    \includegraphics[width=0.5\linewidth]{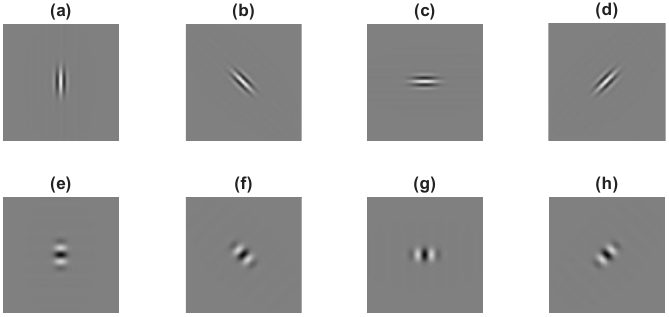}
    \caption{
    Gabor feature kernels with wavelength $\lambda = 2 \times$ probe size, spatial aspect ratio = 1/2, and orientations of 0°, 45°, 90°, and 135°. Subfigures (a), (b), (c), and (d) correspond to each orientation, respectively.
    Gabor probability kernels with wavelength $\lambda = 2 \times$ probe size, spatial aspect ratio = 2, and orientations of 90°, 135°, 180°, and 225°. Subfigures (e), (f), (g), and (h) correspond to each orientation, respectively.}
    \label{fig:kernels}
\end{figure}

To implement this, we utilise the Gabor feature kernel (Fig.~\ref{fig:kernels}(a-d)) and the Gabor probability kernel (Fig.~\ref{fig:kernels}(e-h). The feature kernel is effective for detecting orientation and spatial frequency, while the probability kernel is used to construct the probability map for generating subsequent probes. For example, if the vertical Gabor feature kernel (Fig.~\ref{fig:kernels}(a)) yields a strong response at certain locations, it suggests the presence of structural elements aligned in that direction. Consequently, the corresponding vertical Gabor probability kernel (Fig.~\ref{fig:kernels}(e)) assigns higher probabilities along the same direction in those regions, indicating a greater likelihood of informative content.

\begin{figure*}
\centering
\includegraphics[width=\linewidth]{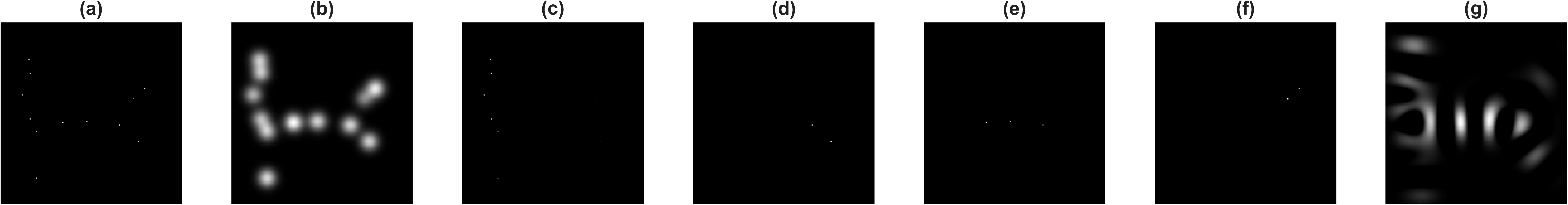}
\caption{This is the middle of the twelfth trial in the first subject's drawing "H". 
(a) is discrete CCA-weighted pixel points (centre of the disc), $I_{\text{pt}}$, we can find 12 points on the image;
(b) is Gaussian smoothing of $I_{\text{pt}}$, $I_{\text{gs}}$, to highlight low-frequency structural information;
(c-f) is a weighted feature map, $I_{\text{feat}}$, for 4 different directions, 0, 45, 90, and 135 degrees. Different points are weighted/highlighted in various directions.
g) is the probability map, $I_{\text{prob}}$.}
\label{fig:Gabor_policyFunc}
\end{figure*}

\subsection{Data-driven Policy} 

\begin{figure}[H]
    \centering
    \includegraphics[width=.5\linewidth]{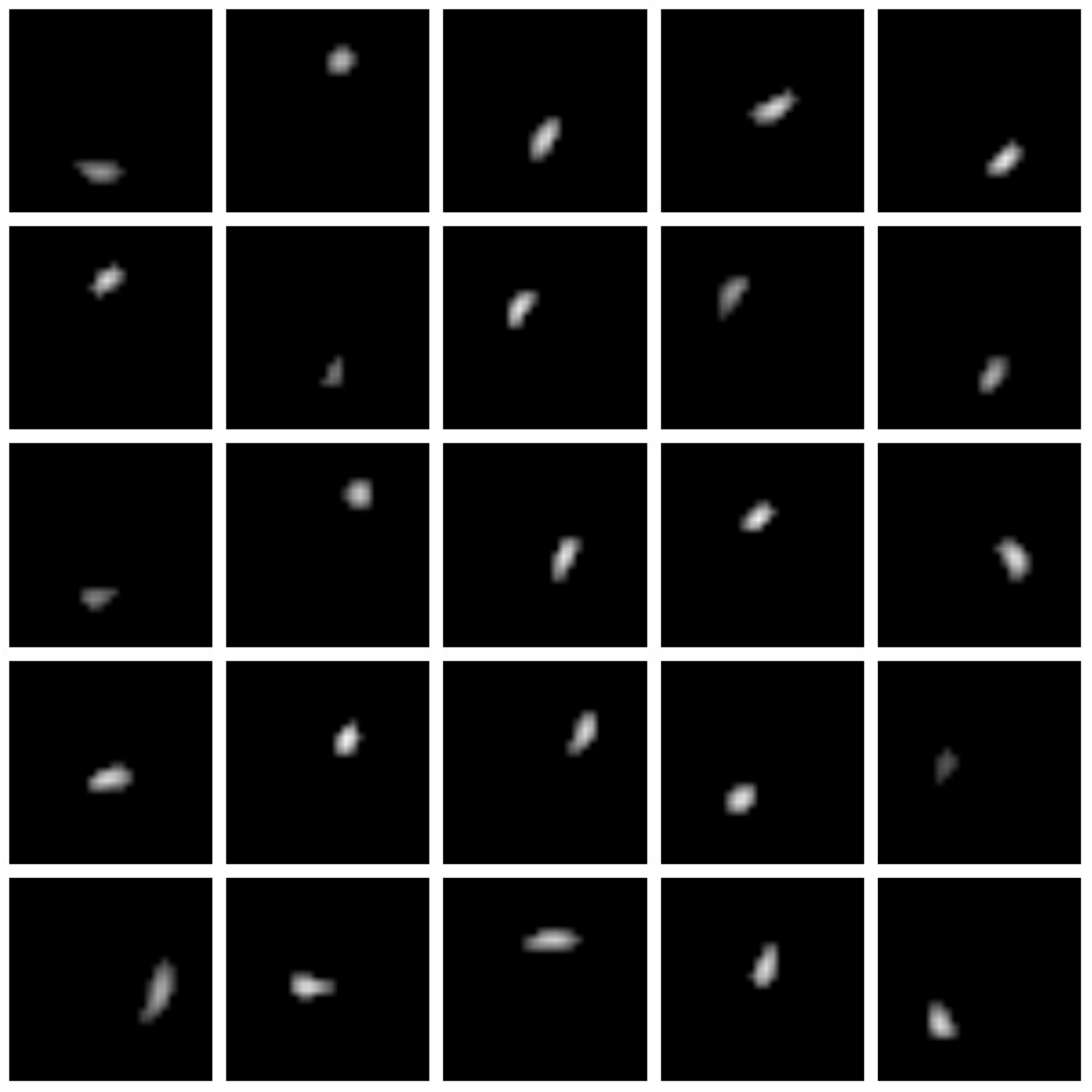}
    \caption{25 atom patterns derived from the MNIST dataset using Non-negative Matrix Factorisation (NNMF). Each atom represents a unique feature extracted from the dataset.}
    \label{fig:atom_patterns}
\end{figure}

Initially, the atom/feature patterns are derived from the MNIST dataset using Non-negative Matrix Factorisation (NNMF). As shown in Fig~\ref{fig:atom_patterns}, we select 25 atoms from the NNMF decomposition to serve as the feature space for image representation and reconstruction.
It is important to note that the original atom patterns derived from NNMF contained 40 feature patterns, and we selected 20 out of 40 feature patterns. Some of which exhibited multiple clustered features. To ensure suitability for our task, we separated these patterns, ensuring that each atom represents a single distinct feature. This preprocessing step is crucial for the subsequent reconstruction task, as it guarantees that each atom effectively captures a unique aspect of the imagined image.

\begin{figure}[H]
    \centering
    \includegraphics[width=0.5\linewidth]{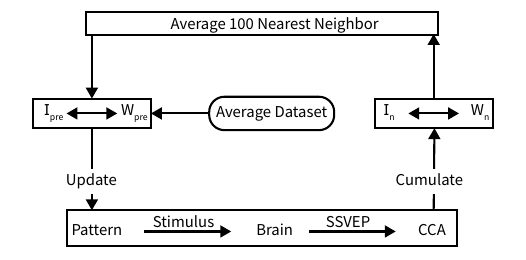}
    \caption{The workflow for Data-driven Policy.}
    \label{fig:DataDriven}
\end{figure}

The workflow for data-driven policy as shown in Fig.~\ref{fig:DataDriven}.
Starting from the avearge dataset image as the $I_{\text{pre}}$, its latent space vector $W_{\text{pre}}$ is used to update the pattern for stimulating the brain. The triggered SSVEP was collected and cumsumed with the corresponding pattern as the reconstructed image, $I_{n}$. In the latent space, the 100 nearest neighbour image was averaged as the new $I_{\text{pre}}$ for the next iteration.

\section{Image Processing and Results} 
\subsection{Binarisation and Registration} 

The reconstructed image is binarised to obtain
$I_{\mathrm{bin}}$ by thresholding the smoothed reconstruction $I_{\text{gs}}$ at
Its mean value:
\begin{equation}
   I_{\text{bin}}(x,y) = 
\begin{cases}
1, & I_{\text{gs}}(x,y) > \overline{I_{\text{gs}}}, \\[6pt]
0, & \text{otherwise}.
\end{cases}
\end{equation}
Here, $I_{\text{gs}}$ denotes the Gaussian-smoothed reconstruction, and $\overline{I_{\text{gs}}}$ is its mean intensity. Pixels with values above the mean are set to 1 (white), and those below are set to 0 (black), yielding a binary image
for subsequent quantitative evaluation.

The subject’s hand-drawn image, serving as ground truth, is spatially aligned with $I_{\mathrm{bin}}$ and denoted as $I_{\text{ref}}$. Specifically, image registration is performed using an affine transformation. The reference image is first resized to match the reconstructed image. Then, an affine (similarity) transformation is estimated to align the reference image to the binarised reconstructed image. The registered reference image is then used for quantitative comparison.

To quantitatively assess reconstruction accuracy, we employ the cosine similarity (COSS), which evaluates the degree of overlap between the binarised reconstructed image and the reference image:
\begin{equation}
    \mathrm{COSS}(I_{\mathrm{bin}}, I_{\mathrm{ref}}) = \frac{\sum_{i=1}^{N}I_{\mathrm{bin}}(i)
    \cdot I_{\mathrm{ref}}(i)}{\sqrt{\sum_{i=1}^{N}I_{\mathrm{bin}}(i)^{2}}
    \cdot \sqrt{\sum_{i=1}^{N}I_{\mathrm{ref}}(i)^{2}}}
\end{equation}
where $I_{\mathrm{bin}}$ and $I_{\mathrm{ref}}$ are the binarised reconstructed and ground truth images, respectively, and $N$ is the total number of pixels.

The binarisation of the predicted image in the Data Driven method, $I_{\text{pre}}$, is defined as follows.  
First, the threshold $T$ is computed from the predicted image $I_{\text{pre}}$ as  
\begin{equation}
    T = \operatorname{mean}(I_{\text{pre}}) 
        + \operatorname{std}\!\left(I_{\text{pre}} \;\middle|\; I_{\text{pre}} > 0\right),
\end{equation}
where $\operatorname{mean}(\cdot)$ denotes the average pixel intensity, and $\operatorname{std}(\cdot)$ denotes the standard deviation.  
Specifically, $\operatorname{std}\!\left(I_{\text{pre}} \;\middle|\; I_{\text{pre}} > 0\right)$ indicates that the standard deviation is calculated only over the set of non-zero pixel values.  

Based on this threshold, the binary image $I_{\text{pre}}^{\text{binary}}$ is obtained as  
\begin{equation}
   I_{\text{pre}}^{\text{binary}}(x,y) = 
\begin{cases}
1, & I_{\text{pre}}(x,y) > T, \\[6pt]
0, & \text{otherwise}.
\end{cases}
\end{equation}
This procedure highlights salient regions in the predicted image while suppressing background noise.


\subsection{MNIST Image Results}
Here we shoe additional MNIST image results i.e. for the MNIST digits "7", "2", "4", and "8".
\begin{figure}[htbp]
    \centering
    \includegraphics[width=0.5\linewidth]{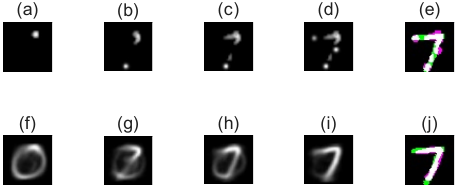}
    \caption{Reconstruction results for the MNIST digit “7.” 
    The first row (a–g) shows the step-by-step reconstruction of the digit from SSVEP signals. The second row (i–o) presents the corresponding predicted images generated by the system at each step. 
    Panels (h) and (p) overlay the binarised reconstructed and binarised predicted images. }
    \label{fig:MNIST_7_reconstruction}    
\end{figure}

\begin{figure}[htbp]
    \centering
    \includegraphics[width=0.5\linewidth]{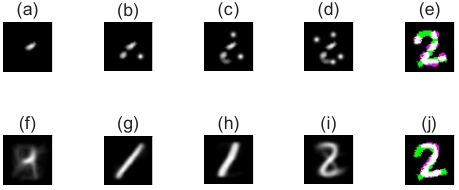}
    \caption{Reconstruction results for the MNIST digit “2.” 
    The first row (a–g) shows the step-by-step reconstruction of the digit from SSVEP signals. 
    The second row (i–o) presents the corresponding predicted images generated by the system at each step. 
    Panels (h) and (p) overlay the binarised reconstructed and binarised predicted images. }
    \label{fig:MNIST_2_reconstruction}    
\end{figure}

\begin{figure}[htbp]
    \centering
    \includegraphics[width=0.5\linewidth]{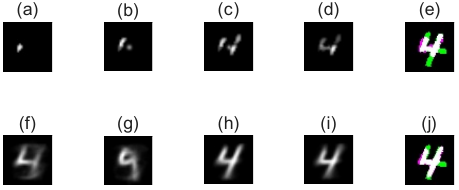}
    \caption{Reconstruction results for the MNIST digit “4.” 
    The first row (a–g) shows the step-by-step reconstruction of the digit from SSVEP signals. 
    The second row (i–o) presents the corresponding predicted images generated by the system at each step. 
    Panels (h) and (p) overlay the binarised reconstructed and binarised predicted images. }
    \label{fig:MNIST_4_reconstruction}    
\end{figure}

\begin{figure}[htbp]
    \centering
    \includegraphics[width=0.5\linewidth]{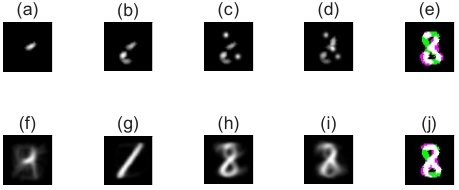}
    \caption{Reconstruction results for the MNIST digit “8.” 
    The first row (a–g) shows the step-by-step reconstruction of the digit from SSVEP signals. 
    The second row (i–o) presents the corresponding predicted images generated by the system at each step. 
    Panels (h) and (p) overlay the binarised reconstructed and binarised predicted images. }
    \label{fig:MNIST_8_reconstruction}    
\end{figure}


\begin{table}[H]
    \centering
    \caption{COSS values for binarised reconstructed images (BRI), 
             binarised predicted images (BPI), and predicted images (PI) 
             across different MNIST digits.}
    \begin{tabular}{lcccc}
        \toprule
        MNIST digit & 7 & 2 & 4 & 8 \\
        \midrule
        BRI & 0.75 & 0.67 & 0.70 & 0.69 \\
        BPI & 0.78 & 0.70 & 0.79 & 0.75 \\
        PI  & 0.82 & 0.78 & 0.82 & 0.82 \\
        \bottomrule
    \end{tabular}
    \label{tab:coss}
\end{table}

\subsection{Stable Diffusion Parameters and Results}
Here we summarise the main parameters and prompts used for the stable diffusion image reconstructions. We also show additional image reconstructions and evolution of the reconstructed images during the iterative mind-drawing procedure.

\textbf{Stable Diffusion checkpoint}\\
v1-5-pruned-emaonly.safetensors [6ce0161689]

\textbf{Prompt for the Robot}

Positive prompt: A highly detailed humanoid robot, full body with head, arms, and legs, in a dynamic pose matching the input sketch.
Futuristic cybernetic design with realistic metallic joints and intricate machinery.
Cinematic lighting, ultra-sharp focus, 8K photorealism, anatomically correct human-like proportions, symmetrical body.
Extremely detailed mechanical textures, polished metal surfaces, realistic reflections, and complex machinery.

Negative prompt: blurry, low quality, cropped, out of frame,
missing limbs, extra limbs, distorted or deformed body parts,
deformed hands, unnatural proportions, asymmetrical body, cartoonish, unrealistic,
low resolution, overexposed, noisy background

\textbf{Prompt for the Tree}

Positive prompt: A masterpiece, ultra-detailed, photorealistic, with a full tree growing from the ground, including visible roots, a branch trunk, and the whole tree is included, visible exposed root system anchored in natural soil with small stones and grass, thick textured trunk with deep bark cracks, knots and patches of moss

Negative prompt: cartoon, illustration, low-res, blurry, watermark, signature, text, over-saturated, pastel, painterly brush strokes, surreal elements, floating or disconnected roots, duplicated or mirrored branches, mutated/deformed shapes, extra limbs, unrealistic glossy plastic look, noisy artifacts, low detail, 3D render/artificial shading

\textbf{Prompt for the desk lamp}

Positive prompt: A full, photorealistic metal desk lamp, with the lamp at the top, a solid base at the bottom, connected by a sleek support arm. The lamp is fully visible from top to bottom, showing detailed textures of metal, reflections, and shadows. High-resolution, ultra-detailed, realistic lighting, studio-quality rendering, showing the complete structure of the lamp, including the base, support, and lampshade

Negative prompt: cartoon, low-res, blurry, text, watermark, cropped, multiple lamps, missing parts, non-metal materials, extra objects

\textbf{Prompt for the aircraft}

Positive prompt: A full, photorealistic metal airplane, with complete wings and tail, wings attached at the midsection of the fuselage. The entire airplane is fully visible in the frame, showing detailed textures of metal, reflections, rivets, and realistic lighting. High-resolution, ultra-detailed, realistic shadows and highlights, studio-quality rendering, complete airplane including fuselage, wings, and tail, viewed from a clear perspective.
Negative prompt: cartoon, illustration, sketch, low-resolution, blurry, cut-off, incomplete, stylized, abstract, unrealistic proportions

\textbf{Other Parameters}

Steps: 20, Sampler: DPM++ 2M, Schedule type: Karras, CFG scale: 7, Seed: 1, Size: 512x512, Model hash: 6ce0161689, Model: v1-5-pruned-emaonly, Denoising strength: 0.75, Version: v1.10.1

\begin{figure}[t]
    \centering
    \includegraphics[width=0.8\linewidth]{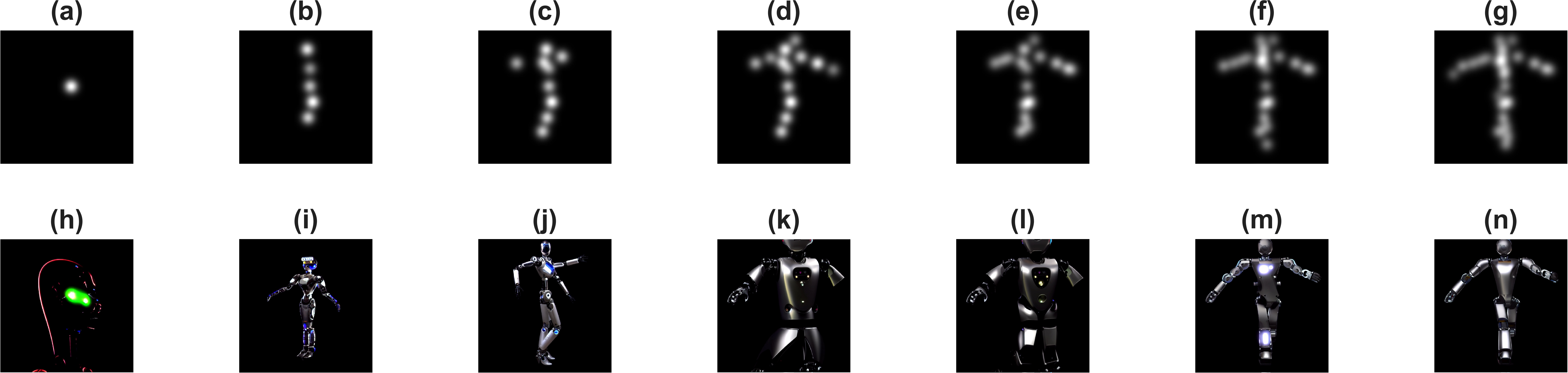}
    \caption{Illustration of the image reconstruction process using Stable Diffusion. Panels~(a–g) present the progressive refinement of mind-drawn sketches  every 4 iteration steps i.e., iteration = 1,5,9..., while panels~(h–n) show the corresponding photorealistic humanoid robot renderings generated by Stable Diffusion, conditioned on the sketch and a robot-related textual prompt.
    }
    \label{fig:SD_example1}
\end{figure}
\begin{figure}[t]
    \centering
    \includegraphics[width=.3\linewidth]{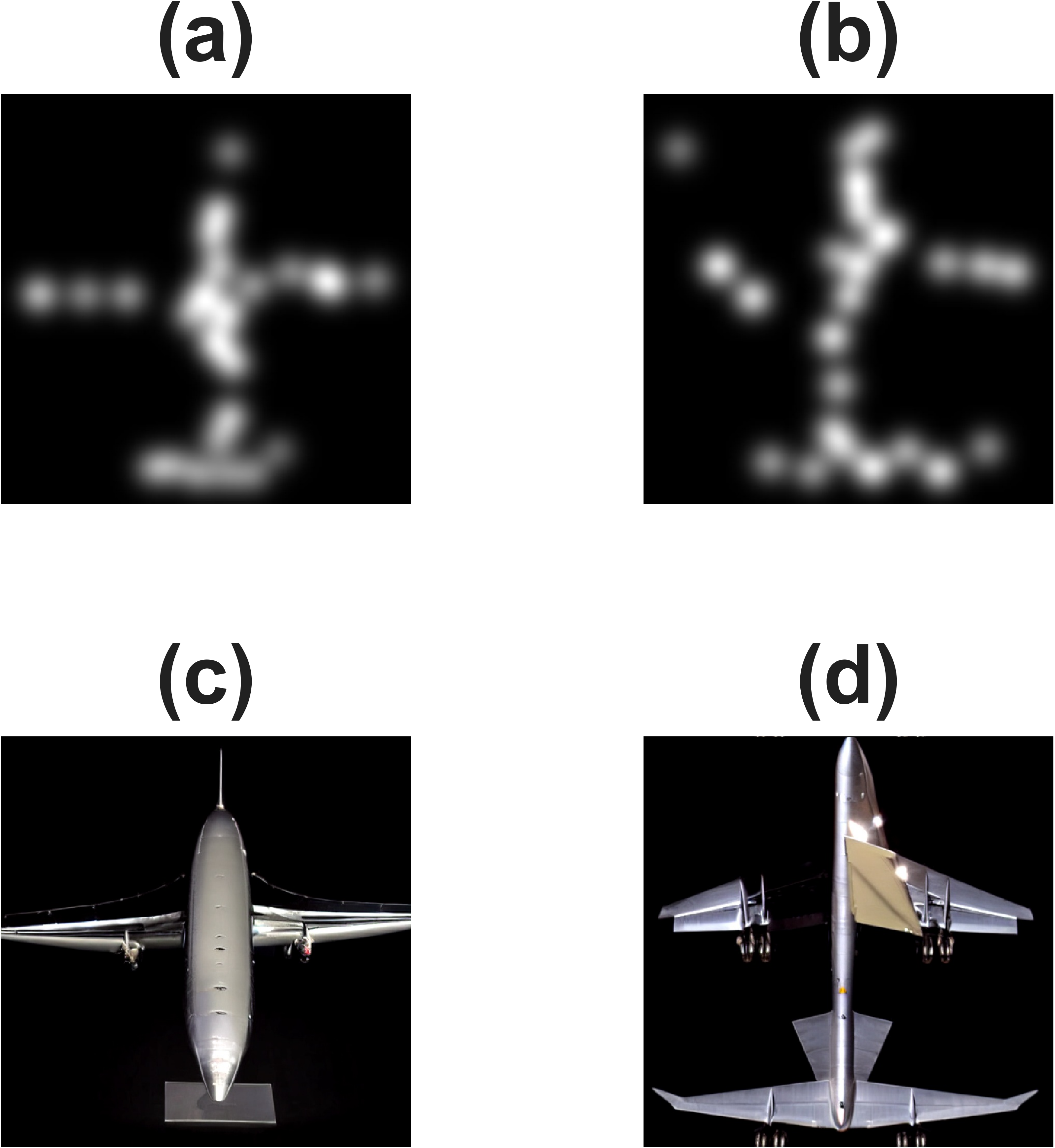}
    \caption{Examples of the reconstruction process with Stable Diffusion. Panels~(a–b) show the same object conceptual sketches from two subjects, whereas panels~(c-d) present the corresponding high-fidelity images generated under the same prompt conditions.}
    \label{fig:SD_example3}
\end{figure}

\newpage
